\documentclass[aps,prb,twocolumn,amsmath,amssymb,showpacs]{revtex4}

\usepackage{graphicx}
\usepackage{dcolumn}
\usepackage{bm}

\begin{document}

\title{Analysis of $f$-$p$ model for octupole ordering in NpO$_2$}

\author{Katsunori Kubo}
\author{Takashi Hotta}

\affiliation{Advanced Science Research Center,
Japan Atomic Energy Research Institute,
Tokai, Ibaraki 319-1195, Japan}

\date{\today}

\begin{abstract}
In order to examine the origin of octupole ordering in NpO$_2$,
we propose a microscopic model constituted of neptunium $5f$ and
oxygen $2p$ orbitals.
To study multipole ordering, we derive effective multipole
interactions from the $f$-$p$ model
by using the fourth-order perturbation theory
in terms of $p$-$f$ hopping integrals.
Analyzing the effective model numerically, we find a tendency
toward $\Gamma_{5u}$ antiferro-octupole ordering.
\end{abstract}



\maketitle




In the research field of condensed matter physics,
it is one of currently important issues
to unveil the mechanism of multipole ordering phenomena
frequently observed in $f$-electron systems.
It has been widely recognized that quadrupole ordering realizes
in several $f$-electron compounds,
but recently, in addition to dipole and quadrupole ordering,
a possibility of ordering of higher order magnetic multipoles,
i.e., octupoles, has been discussed intensively.

A typical candidate with octupole ordering
is the low-temperature ordered phase of NpO$_2$,~\cite{Santini}
since time reversal symmetry is broken
in this phase,~\cite{Dunlap,Kopmann}
but the detected internal field is too weak to be ascribed
to dipole ordering.~\cite{Dunlap,Kopmann,Heaton}
Indeed, several experimental facts~\cite{Westrum,Ross,Mannix,Lovesey}
can be consistently explained
by assuming longitudinal triple-$\mathbf{q}$ $\Gamma_{5u}$
octupole ordering.~\cite{Paixao}
In addition, recent experiments on the $^{17}$O NMR also
support the triple-$\mathbf{q}$ ordered state.~\cite{Tokunaga,Sakai2}

\begin{figure}
\begin{center}
  \includegraphics[width=7cm]{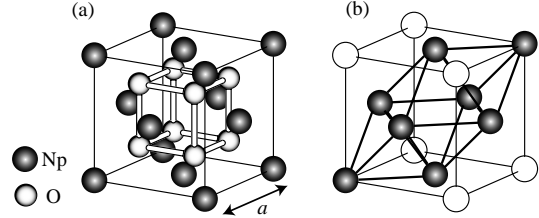}
  \caption{\label{figure:NpO2}
    (a) Crystal structure of NpO$_2$.
    $a$ is the lattice constant.
    (b) 8-site cluster (gray spheres)
    on the fcc lattice taken in our numerical calculation.
  }
\end{center}
\end{figure}

In order to understand why such higher-order multipole
order is realized in NpO$_2$,
it has been highly required to proceed to the microscopic research.
In general, it is difficult to develop a microscopic theory
for complex multipole ordering in $f$-electron systems
beyond the phenomenological level,
but it has been recently proposed to construct a microscopic $f$-electron model
on the basis of a $j$-$j$ coupling scheme.~\cite{Hotta}
Following this proposal, we have studied the $f$-electron model
on an fcc lattice composed of Np ions
with hopping integrals via $(ff\sigma)$ bonding,
and actually found the triple-$\mathbf{q}$ octupole ordering.~\cite{Kubo:NpO2}
However, as shown in Fig.~\ref{figure:NpO2}(a), oxygen anions exist between
Np ions in actual crystal structure.
Thus, it is important to clarify how the octupole ordering
in the $f$-electron model is affected by oxygen anions.

In this paper, we construct a more realistic model
including also $p$ orbitals of oxygen anions
in addition to $f$ orbitals of Np ions.
We derive an effective multipole interaction model by evaluating
the exchange of $f$ electrons via $p$ orbitals
within the fourth-order perturbation with respect to
$p$-$f$ hopping integrals.
By analyzing the effective model, we again find a tendency toward
the triple-$\mathbf{q}$ octupole ordered phase,
indicating that the $f$-electron model on the fcc lattice
have grasped the essential point on the appearance of
octupole ordering of NpO$_2$.

Let us discuss local $f$-electron states of actinide dioxides
based on the $j$-$j$ coupling scheme,
in which we first include the spin-orbit interaction
and consider only the states with total angular momentum $j$=5/2.
From a quantitative viewpoint,
this simplification may not be appropriate for actinide dioxides,
since they are insulators,
and the $LS$ coupling scheme is expected to work well
to describe $f$-electron states in these materials.
However, both schemes are continuously connected to each other
by changing the ratio of the strength of the spin-orbit interaction
and the Coulomb interaction, as long as the symmetry of the ground
state is not changed.~\cite{Hotta:sku}
Thus, on the basis of a spirit of adiabatic continuation,
we expect that qualitative properties at low temperatures
can be grasped whether we choose the $j$-$j$ coupling
or the $LS$ coupling schemes as a starting approximation.
The $j$=5/2 states split into $\Gamma_7$ doublet and $\Gamma_8$
quartet under a cubic crystalline electric field (CEF).
Since the $\Gamma_7$ wavefunction extends along the [111] direction
and an oxygen anion locates in this direction,
the $\Gamma_7$ level is expected to be higher than the $\Gamma_8$ level.
If we assume that the level splitting $\Delta$ between $\Gamma_8$ and
$\Gamma_7$ is large enough, CEF ground states for $f^2$, $f^3$,
and $f^4$ are obtained by accommodating two, three, and four electrons
into $\Gamma_8$ levels,
leading to $\Gamma_5$, $\Gamma_8$, and $\Gamma_1$, respectively,
consistent with experimental results for
UO$_2$,~\cite{Kern2} NpO$_2$,~\cite{Fournier} and PuO$_2$,~\cite{Kern}
respectively.
Thus, we ignore the $\Gamma_7$ state to discuss the ground state of
actinide dioxides in the $j$-$j$ coupling scheme.

We have three comments on the CEF level scheme.
(i) In our picture, the first excited state of PuO$_2$ should
include three $\Gamma_8$ and one $\Gamma_7$ electrons,
indicating that the excitation energy provides
the lower limit for $\Delta$.
Since the CEF excitation energy of PuO$_2$ is 123~meV,~\cite{Kern}
$\Delta$ should be larger than 1400K,
consistent with the initial assumption.
(ii) The $f^3$ state in NpO$_2$ is regarded as the one-hole state
in $\Gamma_8$. In the following, we use a hole picture and
``electron'' denotes such a hole.
(iii) Among the $f^2$ states, the $\Gamma_5$ triplet is the ground
state as observed in UO$_2$. Since the CEF excitation energy in UO$_2$
is as large as 150~meV,~\cite{Kern2} we consider only the $\Gamma_5$
triplet among $f^2$ intermediate states
to study exchange processes of $f^3$ ions in NpO$_2$.


The $\Gamma_8$ quartet consists of two Kramers doublets, and
it is convenient to introduce orbital index $\tau$ ($=\alpha, \beta$)
to label the two Kramers doublets
and spin index $\sigma$ ($=\uparrow, \downarrow$)
to distinguish the two states in each Kramers doublet.
In the second-quantized form,
annihilation operators for $\Gamma_{8}$ electrons are defined by
$f_{\mathbf{r} \alpha \uparrow}$
=$\sqrt{5/6} a_{\mathbf{r} 5/2}+\sqrt{1/6} a_{\mathbf{r} -3/2}$,
$f_{\mathbf{r} \alpha \downarrow}$
=$\sqrt{5/6} a_{\mathbf{r} -5/2}+\sqrt{1/6} a_{\mathbf{r} 3/2}$,
$f_{\mathbf{r} \beta \uparrow}$=$a_{\mathbf{r}  1/2}$,
and
$f_{\mathbf{r} \beta \downarrow}$=$a_{\mathbf{r} -1/2}$,
where $a_{\mathbf{r} j_z}$ is the annihilation operator for an $f$ electron
with the $z$-component $j_z$ of the total angular momentum $j$=5/2
at site $\mathbf{r}$.
Multipole operators are usually expressed as
$X^{\Gamma_{\gamma}}_{\mathbf{r}}$,
where $\Gamma_{\gamma}$ denotes symmetry.
The explicit forms of $X^{\Gamma_{\gamma}}_{\mathbf{r}}$
in the $\Gamma_8$ subspace are found in Ref.~\onlinecite{Kubo:NpO2}.


Now we show the $f$-$p$ model for NpO$_2$, given by
\begin{equation}
  \mathcal{H}
  =\mathcal{H}_{f}+\mathcal{H}_{p}
  +\mathcal{H}_{\mathrm{kin}},
  \label{eq:model}
\end{equation}
where $\mathcal{H}_{f}$ and $\mathcal{H}_{p}$ are the local
$f$- and $p$-electron terms, respectively, and
$\mathcal{H}_{\mathrm{kin}}$ denotes the hybridization term
between $f$ and $p$ electrons.
Among them, the local $f$-electron term is explicitly given by
\begin{equation}
  \begin{split}
    \mathcal{H}_{f}
    =&\epsilon_f
    \sum_{\mathbf{r},\tau}
    n_{\mathbf{r} \tau}
    +U\sum_{\mathbf{r} \tau}
    n_{\mathbf{r} \tau \uparrow} n_{\mathbf{r} \tau \downarrow}
    +U^{\prime}\sum_{\mathbf{r}} n_{\mathbf{r} \alpha} n_{\mathbf{r} \beta}
    \\
    &+J\sum_{\mathbf{r},\sigma,\sigma^{\prime}}
    f^{\dagger}_{\mathbf{r} \alpha \sigma}
    f^{\dagger}_{\mathbf{r} \beta \sigma^{\prime}}
    f_{\mathbf{r} \alpha \sigma^{\prime}}
    f_{\mathbf{r} \beta \sigma}
    \\
    &+J^{\prime}\sum_{\mathbf{r},\tau \ne \tau^{\prime}}
    f^{\dagger}_{\mathbf{r} \tau \uparrow}
    f^{\dagger}_{\mathbf{r} \tau \downarrow}
    f_{\mathbf{r} \tau^{\prime} \downarrow}
    f_{\mathbf{r} \tau^{\prime} \uparrow},
  \end{split}
  \label{f_local}
\end{equation}
where $\epsilon_f$ denotes the energy level of $\Gamma_8$,
$n_{\mathbf{r} \tau \sigma}
=f^{\dagger}_{\mathbf{r} \tau \sigma} f_{\mathbf{r} \tau \sigma}$,
and $n_{\mathbf{r} \tau}$=$\sum_{\sigma} n_{\mathbf{r} \tau \sigma}$.
The coupling constants $U$, $U^{\prime}$, $J$, and $J^{\prime}$
denote the intra-orbital Coulomb, inter-orbital Coulomb, exchange,
and pair-hopping interactions, respectively.

The local $p$-electron term is expressed as
\begin{equation}
  \begin{split}
    \mathcal{H}_{p}
    =&\epsilon_p \sum_{\mathbf{x} m,\sigma}
    p^{\dagger}_{\mathbf{x} m \sigma} p_{\mathbf{x} m \sigma}\\
    &+\text{(on-site interactions for $p$ orbitals)},
  \end{split}
\end{equation}
where $\epsilon_p$ denotes the energy level of $p$ orbitals of oxygen ions
and $p_{\mathbf{x} m \sigma}$ is the annihilation operator
for an electron with the $z$-component $m$ of orbital angular momentum
and real spin $\sigma$ at O site $\mathbf{x}$.
Finally, $\mathcal{H}_{\mathrm{kin}}$ is given by
\begin{equation}
  \mathcal{H}_{\mathrm{kin}}
  =\sum_{\mathbf{r},m,\sigma, s,\bm{\nu}}
  (t^{\bm{\nu}}_{m \sigma; s}
  p^{\dagger}_{\mathbf{r}+\bm{\nu} m \sigma} f_{\mathbf{r} s} +\mathrm{h.c.}),
\end{equation}
where $\bm{\nu}$ is a vector connecting nearest-neighbor Np and O sites
and $s$ symbolically denotes both spin and orbital of an $f$ electron.
The hopping integrals $t^{\bm{\nu}}_{m \sigma; s}$
are represented in terms of $(pf\sigma)$ and $(pf\pi)$
by using the Slater-Koster table.~\cite{Takegahara}

\begin{figure}
  \includegraphics[width=8cm]{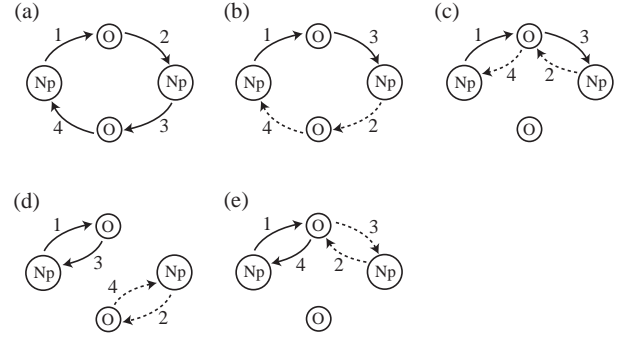}
  \caption{\label{figure:fourth_order_processes}
    Fourth-order processes with respect to $p$-$f$ hoppings.
    (a) Electrons doubly occupy a Np site.
    Electrons occupy (b) two or (c) one O site simultaneously,
    and then exchange.
    Electrons occupy (d) two or (e) one O site simultaneously,
    and then return to the initial sites.
    The solid and dashed lines in (b)--(e) distinguish
    paths for different electrons.
  }
\end{figure}

From the $f$-$p$ model (\ref{eq:model}),
we derive an effective interaction
by using the fourth-order perturbation theory
with respect to the $p$-$f$ hopping integrals.
Here we note that the fourth-order processes are classified into two types:
(a) Two electrons doubly occupy a Np ion in the intermediate states,
as schematically shown in Fig.~\ref{figure:fourth_order_processes}(a).
(b)--(e) Two electrons occupy one or two oxygen sites simultaneously
in the intermediate states,
as shown in figures~\ref{figure:fourth_order_processes}(b)--(e).
The effective Hamiltonian is arranged in the form of
\begin{equation}
  \mathcal{H}_{\rm eff}
  =\sum_{\langle \mathbf{r},\mathbf{r}^{\prime} \rangle, s_1\textrm{--}s_4 }
  I^{\mathbf{r}^{\prime}-\mathbf{r}}_{s_3, s_4; s_1, s_2}
  f^{\dagger}_{\mathbf{r} s_3}
  f_{\mathbf{r} s_1}
  f^{\dagger}_{\mathbf{r}^{\prime} s_4}
  f_{\mathbf{r}^{\prime} s_2},
\end{equation}
where $\langle \mathbf{r}, \mathbf{r}^\prime \rangle$ denotes the pair of
nearest-neighbor sites and the effective interaction $I$ is given by
\begin{equation}
   I^{\bm{\mu}}_{s_3, s_4; s_1, s_2}
  = \sum_{\mathrm{k}=\mathrm{a}\textrm{--}\mathrm{e}}
    I^{\mathrm{(k)} \ \bm{\mu}}_{s_3, s_4; s_1, s_2}.
\end{equation}
In the following, we will briefly explain each term included in $I$.

In order to derive multipole interaction $I^{(a)}$ from the process (a),
we simply replace $f$-electron hopping $t^{\bm{\mu}}_{s; s^{\prime}}$
in the effective interaction $I$ of the $f$-electron model~\cite{Kubo:NpO2}
with the effective $f$-$f$ hopping $T^{\bm{\mu}}_{s; s^{\prime}}$
via $p$ orbitals.
For the purpose, we define $f$-electron hopping
between sites $\mathbf{r}$ and $\mathbf{r}+\bm{\mu}$ via $p$ orbitals of the oxygen ion
at $\mathbf{r}+\bm{\nu}$, given by
\begin{equation}
  T^{\bm{\mu},\bm{\nu}}_{s; s^{\prime}}
  =(\epsilon_f-\epsilon_p)^{-1}\sum_{m \sigma}
  (t^{\bm{\nu}-\bm{\mu}}_{m \sigma; s})^*
  t^{\bm{\nu}}_{m \sigma; s^{\prime}}.
\end{equation}
Then, we obtain
$T^{\bm{\mu}}_{s; s^{\prime}}$=$\sum_{\bm{\nu}} T^{\bm{\mu},\bm{\nu}}_{s; s^{\prime}}$,
since the effective $f$-electron hopping
consists of two paths through different oxygen sites.
For instance, along the [110] direction,
the effective $f$-$f$ hopping integral is given by
$T^{(a/2,a/2,0)}_{s; s^{\prime}}$=
$T^{(a/2,a/2,0),(a/4,a/4, a/4)}_{s; s^{\prime}}$+
$T^{(a/2,a/2,0),(a/4,a/4,-a/4)}_{s; s^{\prime}}$,
where $a$ is the lattice constant.
Using $T^{\bm{\mu}}_{s; s^{\prime}}$, we express $I^{\rm (a)}$ as
\begin{equation}
  \begin{split}
    I^{\mathrm{(a)} \ \bm{\mu}}_{s_3, s_4; s_1, s_2}
    =&-(U^{\prime}-J)^{-1}
    \\
    \times
    \sum_{u,s,s^{\prime}}
    &\left[(T^{\mathbf{r}^{\prime}-\mathbf{r}}_{s^{\prime}; s_4})^*
      P^*_{u; s_3, s^{\prime}}
      P_{u; s_1, s}
      T^{\mathbf{r}^{\prime}-\mathbf{r}}_{s; s_2}\right.
      \\
      &+
      \left.
      (T^{\mathbf{r}-\mathbf{r}^{\prime}}_{s^{\prime}; s_3})^*
      P^*_{u; s_4, s^{\prime}}
      P_{u; s_2, s}
      T^{\mathbf{r}-\mathbf{r}^{\prime}}_{s; s_1}\right],
  \end{split}
\end{equation}
where $P_{u; s, s^{\prime}}$ denotes
the inner product between
one of the $\Gamma_5$ triplet states denoted by $u$ and
the $f^2$ state labeled by $s$ and $s^{\prime}$.

We note that the effective $f$-$f$ hopping integrals have the same form
as those via $(ff\sigma)$ bonding,~\cite{Kubo:NpO2}
for instance,
\begin{equation}
  T^{(a/2,a/2,0)}_{\tau \sigma; \tau^{\prime} \sigma^{\prime}} \propto
  [\delta_{\tau \tau^{\prime}} \delta_{\sigma \sigma^{\prime}}
    +c_1 \sigma^z_{\tau \tau^{\prime}} \delta_{\sigma \sigma^{\prime}}
    +c_2 \sigma^y_{\tau \tau^{\prime}} \sigma^z_{\sigma \sigma^{\prime}}],
\end{equation}
where $\bm{\sigma}$ are Pauli matrices, and
$c_1$ and $c_2$ are constants depending on $(pf\sigma)$ and $(pf\pi)$.
This fact indicates that the form of the hopping integrals are restricted
by $f$-electron symmetry, and the simple $(ff\sigma)$ model may grasp
properties of actual materials with complex structures.


Concerning processes (b)--(e),
we consider the effect of the Coulomb interaction at oxygen sites,
symbolically expressed as ``$U_p$''.
In this paper, we study two limiting cases,
$U_p$=0 and $U_p$=$\infty$.
For $U_p$=0, multipole interaction contains all the processes (b)--(e),
given by
\begin{align}
  I^{\mathrm{(b)} \ \bm{\mu}}_{s_3, s_4; s_1, s_2}
  &=2(\epsilon_p-\epsilon_f)^{-1}
  \sum_{\bm{\nu}}
  T^{\bm{\mu}, \bm{\nu}}_{s_4; s_1}
  T^{-\bm{\mu},-\bm{\nu}}_{s_3; s_2},
  \\
  I^{\mathrm{(c)} \ \bm{\mu}}_{s_3, s_4; s_1, s_2}
  &=2(\epsilon_p-\epsilon_f)^{-1}
  \sum_{\bm{\nu}}
  T^{\bm{\mu}, \bm{\nu}}_{s_4; s_1}
  T^{-\bm{\mu},\bm{\nu}-\bm{\mu}}_{s_3; s_2},
  \\
  I^{\mathrm{(d)} \ \bm{\mu}}_{s_3, s_4; s_1, s_2}
  &=2(\epsilon_p-\epsilon_f)^{-1}
  \sum_{\bm{\nu}}
  S^{\bm{\nu}}_{s_4; s_1}
  S^{-\bm{\mu}+\bm{\nu}}_{s_3; s_2},
\end{align}
and
\begin{equation}  
  I^{\mathrm{(e)} \ \bm{\mu}}_{s_3, s_4; s_1, s_2}
  =-I^{\mathrm{(d)} \ \bm{\mu}}_{s_3, s_4; s_1, s_2},
\end{equation}
respectively, where
\begin{equation}
  S^{\bm{\nu}}_{s; s^{\prime}}
  =(\epsilon_f-\epsilon_p)^{-1}
  \sum_{m \sigma}
  (t^{\bm{\nu}}_{m \sigma; s})^*
  t^{\bm{\nu}}_{m \sigma; s^{\prime}}.
\end{equation}
Note that the sum of processes (d) and (e) merely becomes
energy shift, since
$I^{\mathrm{(e)} \ \bm{\mu}}_{s_3, s_4; s_1, s_2}
+I^{\mathrm{(d)} \ \bm{\mu}}_{s_3, s_4; s_1, s_2}
\propto \delta_{s_1 s_3} \delta_{s_2 s_4}$,
and such terms can be eliminated in the present discussion.
After all, the multipole interaction for $U_p$=0 is given by
\begin{equation}
  I^{\bm{\mu}}_{s_3, s_4; s_1, s_2}
  = I^{\mathrm{(a)} \ \bm{\mu}}_{s_3, s_4; s_1, s_2}
  + 2(\epsilon_p-\epsilon_f)^{-1} T^{\bm{\mu}}_{s_4; s_1}
  T^{-\bm{\mu}}_{s_3; s_2}.
\end{equation}
Note that this term is expressed by only
the effective $f$-$f$ hopping integral.

\begin{figure}[t]
  \includegraphics[width=7.5cm]{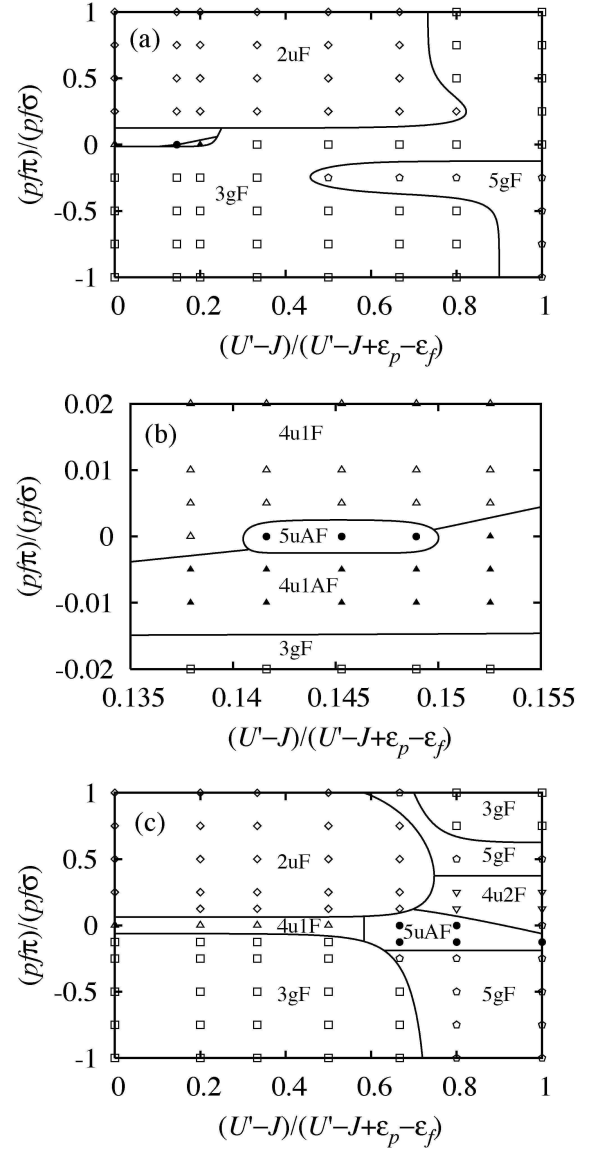}
  \caption{\label{figure:PD_Up0_Upinfty}
    Phase diagrams obtained from the multipole correlation functions.
    Here 3gF denotes $\Gamma_{3g}$ moment with $\mathbf{q}=(0,0,0)$,
    and so on.
    The antiferro (AF) ordering vector is $\mathbf{q}=(0,0,1)$
    in units of $2\pi/a$ for all the AF phases.
    (a) Phase diagram for $U_p$=0. 
    (b) Phase diagram for $U_p$=0 in a magnified scale around
    $(U^{\prime}-J)/(\epsilon_p-\epsilon_f+U^{\prime}-J)$=0.145
    and $(pf \pi)$=0.
    (c) Phase diagram for $U_p$=$\infty$. 
  }
\end{figure}


Now we move on to another limiting case $U_p$=$\infty$,
in which the effect of $U_p$ is included by
prohibiting processes (c) and (e).
Thus, the effective interaction for $U_p$=$\infty$ is given by
\begin{equation}
  I^{\bm{\mu}}_{s_3, s_4; s_1, s_2}
 =I^{\mathrm{(a)} \ \bm{\mu}}_{s_3, s_4; s_1, s_2}
 +I^{\mathrm{(b)} \ \bm{\mu}}_{s_3, s_4; s_1, s_2}
 +I^{\mathrm{(d)} \ \bm{\mu}}_{s_3, s_4; s_1, s_2}.
\end{equation}
It should be noted that in the process (b), a couple of electrons
exchange their sites by avoiding the effect of Coulomb interactions.
Such a term is characteristic to the crystal structure of NpO$_2$,
and it has a tendency to stabilize the octupole ordering,
as shown later.


Using the effective Hamiltonian,
we evaluate numerically the multipole correlation function,
$\chi^{\Gamma_{\gamma}}_{\mathbf{q}}=
(1/N)\sum_{\mathbf{r}, \mathbf{r}^{\prime}}
e^{i \mathbf{q} \cdot (\mathbf{r}-\mathbf{r}^{\prime})}
\langle X^{\Gamma_{\gamma}}_{\mathbf{r}}
X^{\Gamma_{\gamma}}_{\mathbf{r}^{\prime}} \rangle$,
where $\langle \cdots \rangle$ denotes the expectation value
using the ground-state wavefunction.
Here we take $N$=8, as shown in Fig.~\ref{figure:NpO2}(b).
Figures~\ref{figure:PD_Up0_Upinfty}(a) and (c) show phase diagrams,
presenting the multipole moment which has the largest value in
the correlation function at each parameter set,
for $U_p$=0 and $U_p$=$\infty$, respectively.
For $U_p$=0, as shown in Fig.~\ref{figure:PD_Up0_Upinfty}(b)
in a magnified scale, there is a very small, but finite region of
the $\Gamma_{5u}$ antiferro-octupole
[$\mathbf{q}=(0,0,1)$ in units of $2\pi/a$, 5uAF] phase.
For $U_p$=$\infty$, on the other hand,
the region of the 5uAF phase becomes much larger than that for $U_p$=0.
The 5uAF phase locates in the parameter region with
$\epsilon_p-\epsilon_f \ll U'-J$, in which
processes~(b) and (d) work effectively.
Since process~(d) provides only a quadrupole interaction,
we conclude that
the stabilization of the 5uAF phase originates from the process (b).

Note that the Coulomb energy in the $f^2$ intermediate state
$U^{\prime}-J$ is expected to be in the order of 1~eV,
but we cannot estimate the difference of the energy levels
of $p$ and $f$ electrons $\epsilon_p-\epsilon_f$
within the present theory.
We also note that in the phase diagrams,
the 5uAF phase appears for $(pf\pi) \simeq 0$.
Since $(pf\sigma)$ and $(pf\pi)$ are treated as parameters
in this paper, we have no clear answer why the absolute value
of $(pf\pi)$ should be so small for the appearance of the 5uAF phase.
In order to confirm the reality of the parameter region for the
octupole phase obtained in this study is realistic or not,
it is highly requested to perform the band-structure calculations
for NpO$_2$.
This is one of future problems.


In summary, on the basis of the $f$-$p$ model for NpO$_2$,
we have found a finite region of the 5uAF phase
for both cases of $U_p$=0 and $U_p$=$\infty$,
Thus, we expect that this property is retained in the actual situation
for NpO$_2$ with finite $U_p$.
While the ordered state cannot be entirely determined within
the present small-cluster calculation,
it is confirmed that among structures with $\mathbf{q}$=$(0,0,1)$ and
equivalent ones,
the triple-$\mathbf{q}$ structure is energetically favorable,~\cite{Kubo:NpO2}
since the $\Gamma_{5u}$ moment in the $\Gamma_8$ subspace has
the easy axis along the [111] direction.
We emphasize that the $\Gamma_{5u}$ antiferro-octupole phase is
also realized in the simple $(ff\sigma)$ hopping model
on an fcc lattice even without oxygen $p$ orbitals.~\cite{Kubo:NpO2}
These findings indicate that
the tendency toward $\Gamma_{5u}$ antiferro-octupole ordering
is common to $\Gamma_8$ models on fcc lattices.


We thank S. Kambe, N. Metoki, H. Onishi, Y. Tokunaga, K. Ueda,
R. E. Walstedt, and H. Yasuoka for useful discussions.
One of the authors (K. K.) is supported
by the REIMEI Research Resources of Japan Atomic Energy
Research Institute.
Another author (T. H.) is supported by Japan Society for
the Promotion of Science and
by the Ministry of Education, Culture, Sports, Science,
and Technology of Japan.


\end{document}